\documentclass[a4paper]{article}

\usepackage{icrc2013}
\usepackage[english]{babel}

\title{CosMO -- A Cosmic Muon Observer Experiment for Students}

\shorttitle{CosMO -- A Cosmic Muon Observer Experiment for Students}

\authors{
R.~Franke,
M.~Holler,
B.~Kaminsky,
T.~Karg,
H.~Prokoph,
A.~Sch\"onwald,
C.~Schwerdt,
A.~St\"o{\ss}l,
M.~Walter
}

\afiliations{
DESY, Platanenallee 6, 15738 Zeuthen, Germany
}

\email{carolin.schwerdt@desy.de}

\abstract{What are cosmic particles and where do they come from? These
  are questions which are not only fascinating for scientists in
  astrophysics. With the CosMO experiment (Cosmic Muon Observer)
  students can autonomously study these particles. They can perform
  their own hands-on experiments to become familiar with modern
  scientific working methods and to obtain a direct insight into
  astroparticle physics. In this contribution we present the
  experimental setup and possible measurements. The detector consists
  of three scintillator boxes. Events are triggered and readout by a
  data acquisition board developed for the QuarkNet Project. With a
  Python program running on a netbook under Linux, the trigger and
  data taking conditions can be defined. The program displays the
  particle rates in real-time and stores the data for offline
  analysis.  Possible student experiments are the measurement of
  cosmic particle rates dependent on the zenith angle, the
  distribution of geometrical size of particle showers, and the
  lifetime of muons. Twenty CosMO detectors have been built at
  DESY. They are used within the German outreach network
  \textit{Netzwerk Teilchenwelt} at 15 astroparticle-research
  institutes and universities for project work with students.}


\begin{document}
\maketitle

\section{Introduction}

Cosmic particles of various types reach the Earth. They rain down
constantly, some of them with energies much higher than the LHC
reaches. Cosmic rays contribute to natural background radiation and
produce the beautiful light of the auroras. Perhaps they also
influence the formation of clouds and even the evolution of
live. Although hundreds of these particles pass through us every
second, most people do not know about this. Within \textit{Netzwerk
  Teilchenwelt} \cite{Hawner:2012aa, NTW:URL} we have developed the
CosMO experiment to provide insights into this fascinating field.

CosMO is a scintillation counter experiment based on detector
components that are used in particle and astroparticle physics.  It
can be operated by students on their own and brings the current topic
of astroparticle physics to pupils who are interested in physics,
astronomy, or computing. The project allows autonomous investigation
and gets students involved in research. They are given the opportunity
to experience hands-on science with the help of modern
measurement techniques, as well as analysis methods in particle
physics and in close collaboration with scientists. CosMO can be used
in outreach projects at research institutes or within school
teaching. Teachers receive training so that they are enabled to
incorporate the CosMO experiment into their classes.

DESY and other partner institutes within \textit{Netzwerk
  Teilchenwelt} lend the CosMO experiments for student projects and
provide advise and support.

\section{Detector Setup}

\begin{figure*}[t]
  \centering
  \includegraphics[width=\textwidth]{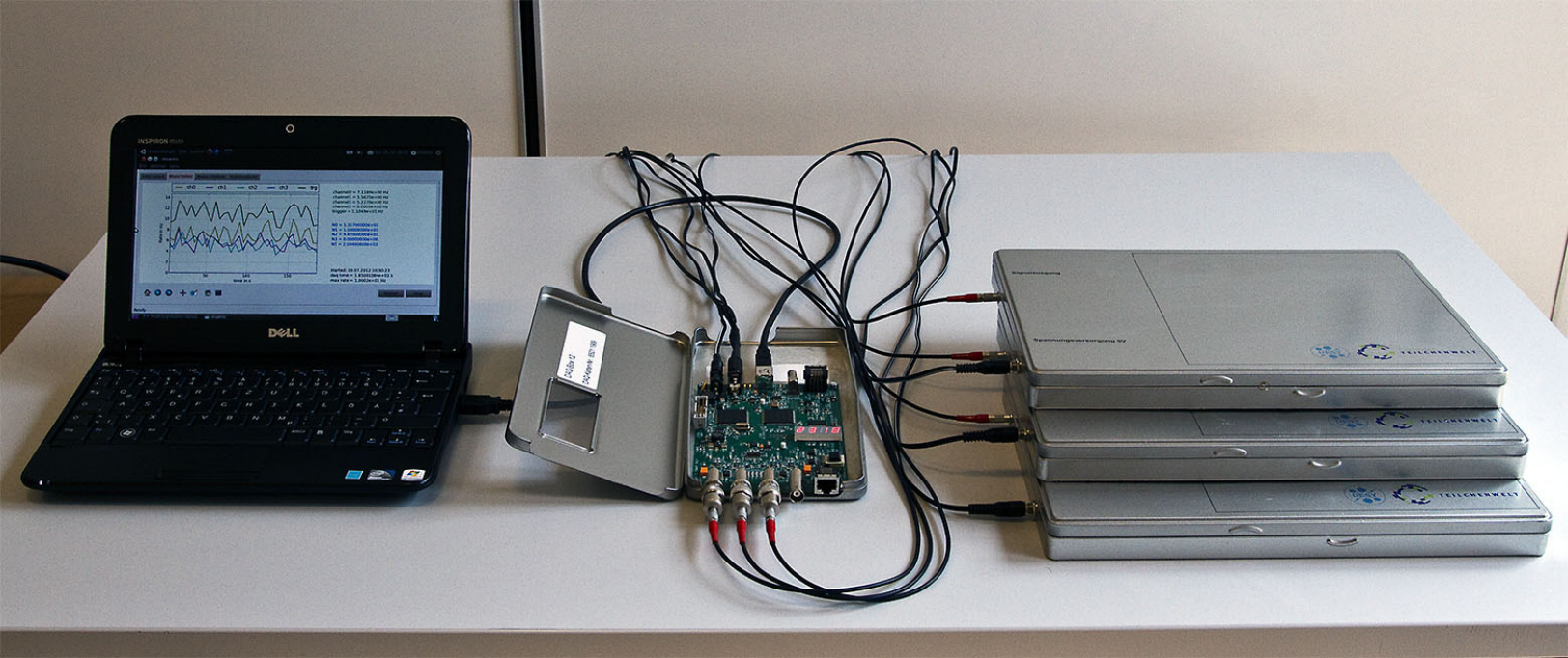}
  \caption{Overview of the CosMO setup in operation with three
    scintillator boxes (right), the DAQ card (center), and the readout
    netbook with the graphical user interface (left).}
  \label{fig:overview}
\end{figure*}

The design goal of the CosMO detector was to develop an
astroparticle-physics experiment that can be operated by students and
that is easily transportable to be used at schools. The detector
consists of three plastic scintillators. The scintillator
tiles are connected to a data acquisition (DAQ) card with
software-adjustable thresholds and coincidence conditions. A netbook
computer is used to control the DAQ readout and to visualize the
data. Figure~\ref{fig:overview} shows the full setup during operation.

\paragraph{Scintillator.} We use plastic scintillator\footnote{Ejen
  Technology EJ-200} tiles with a size of {$20 \times 20 \times
  1.2$~cm$^3$. The tiles are read out with 9 optical fibres each,
  which are connected to a multi-pixel photon
  counter\footnote{Hamamatsu S10931-050P} (MPPC), a type of silicon
  photomultiplier. The operating voltage of about $70$~V for the MPPC
  is generated from a $5$~V input voltage using an adjustable DC-DC
  converter\footnote{iseg APn 02 255 5}. The MPPC has the advantage
  that only low voltages are required for operation which are
  considered safe for students. The scintillator material and MPPC are
  housed in a lightproof aluminium box with connectors for the voltage
  supply and coaxial cable for the analogue MPPC output signal. The
  layout of the scintillator boxes is shown in
  Figure~\ref{fig:scintillator}.

\begin{figure}[t]
  \centering
  \includegraphics[width=0.48\textwidth]{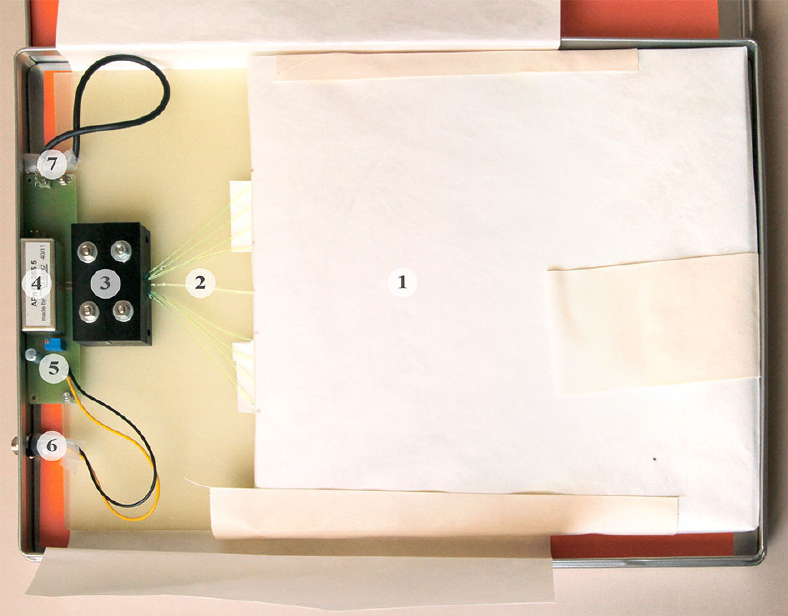}
  \caption{Overview of an open scintillator box: 1.~plastic
    scintillator wrapped in paper, 2.~optical fibers, 3.~MPPC
    mounting, 4.~voltage converter, 5.~resistor for voltage
    adjustment, 6.~connector for power supply, 7.~analogue output for
    MPPC signal.}
  \label{fig:scintillator}
\end{figure}

\paragraph{DAQ card.} The DAQ card used is the version 2.5 data
acquisition board \cite{Hansen:2003aa} developed by a team of
Fermilab, University of Nebraska, and University of Washington for the
\textit{Cosmic Ray e-Lab} \cite{CReLab:URL} of the \textit{QuarkNet}
\cite{QuarkNet:URL} project. Up to four analogue input signals are
processed by discriminators with a software-adjustable
thresholds. Each crossing of the discriminator threshold is

\begin{itemize}
\item recorded by a time-to-digital converter with a resolution of
  $1.25$~ns (rising and falling edge),
\item processed in a complex programmable logic device (CPLD) which
  forms a multiplicity trigger decision,
\item counted in scalers for each channel.
\end{itemize}

The required multiplicity and the trigger time window can both be set
in software. If the trigger condition is fulfilled an event is
generated that contains a global timestamp and all threshold crossing
times. In addition, a GPS module can be connected to the DAQ card that
allows recording the global timestamp in UTC with a precision of
$50$~ns \cite{Berns:2003aa}.

The DAQ card is powered with $5$~V DC and offers the possibility to
distribute its input voltage to the scintillator boxes so that the
whole setup can be operated with a single main power supply.

\paragraph{Readout.} The DAQ card provides a virtual serial port via a
standard USB interface that makes it possible to access the data from
a variety of different operating systems and programming languages. We
developed software (cf.~Sec.~\ref{sec:analysis}) specifically tailored
to the CosMO project and running on a lightweight netbook computer.

\bigskip The complete CosMO setup has a weight of about $5$~kg and can
be transported in a single briefcase, making it well usable in
lectures, labs and outreach projects as well as in school
projects. The detector can also be operated with a small $5$~V battery
pack, independent of the power grid.

\section{Data Processing and Visualisation}
\label{sec:analysis}

Once connected with the netbook, DAQ communications are managed via a
simple text protocol. The discriminator threshold crossing information
is transmitted together with GPS and timing information via the USB
interface. Commands can be issued to the DAQ card via the same
interface.

To manage DAQ communications and visualize the data, we developed the
software \textit{muonic} \cite{muonic:URL}. As the main purpose of
CosMO is the use for student experiments, the focus during the
development of the software was to provide an easy-to-use interface.
\textit{Muonic} was developed entirely in Python, using PyQt4 for the
graphical user interface (GUI). The software is written in a modular
way, and can be extended easily. It is open-source and does not depend
on closed source libraries. \textit{Muonic} runs platform-independent
and does not need an internet connection, which allows its operation
in remote locations.

With \textit{muonic} the user can set the DAQ configuration, like
thresholds or trigger conditions by manipulating typical GUI
elements. The software queries the DAQ for scaler information in a
given time interval and displays this data as a simple rate per
channel over time plot. Also the mean rate is calculated. The width of
PMT pulses is displayed in a histogram for debugging and
maintenance. The data stream from the DAQ can be stored in its raw
format or as tab-separated values that can be imported into a
spreadsheet. A screen shot of the GUI during a typical rate
measurement is shown in Fig.~\ref{fig:gui}.

\begin{figure*}[ht!]
  \centering
  \includegraphics[width=0.9\textwidth]{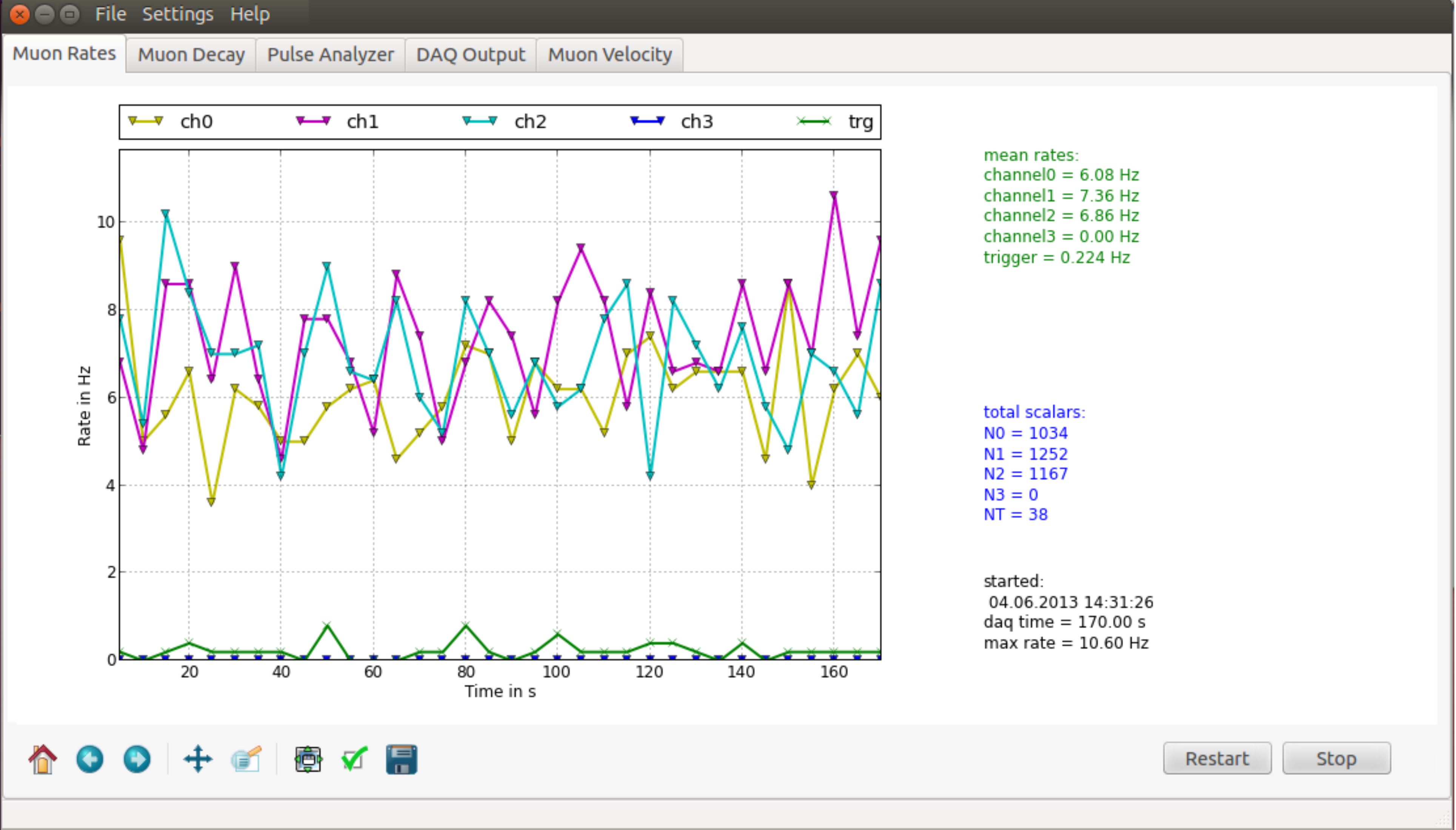}
  \caption{Screen shot of the \textit{muonic} graphical user interface
    during a typical rate measurement.}
  \label{fig:gui}
\end{figure*}

\section{Projects for Students}

\subsection{Calibration}

With \textit{muonic} students can perform calibration measurements of
the scintillation detectors themselves and thus gain deeper
understanding of the physical principles involved. Evaluating the
trigger rate vs.~averaging time students can explore the influence of
statistical fluctuations and determine the measurement time necessary
to get a robust rate estimate. Analysing the single channel trigger
rate vs.~threshold will reveal the desired plateau of stable rates due
to cosmic muons, bound by electronic noise towards too low thresholds
and the absence of muon signals at too high thresholds. A typical
calibration measurement is shown in Fig.~\ref{fig:calibration}. For
each threshold setting approx.~$5$ minutes of data were collected.

\subsection{Zenith Angle Dependence}

A simple project for students to become familiar with the CosMO
experiment is the measurement of the rate of coincidences as a
function of the zenith angle, i.e.~the suppression of the atmospheric
muon flux with increasing atmospheric depth. At least two
scintillation detectors are used in coincidence to suppress electronic
noise and are placed in parallel to each other at a mutual distance
$d$, which defines the solid angle from which particles are
accepted. It has been shown that distances between $20$ and $30$~cm
give a good compromise between restriction to direction of arrival and
too low rates. The coincidence rate is measured using the
\textit{muonic} software and the measurement is repeated at different
zenith angles. The students can be tasked to construct a setup that
allows the rotation to well defined zenith angles while keeping the
scintillator plates parallel to each other and at constant
distance. The resulting rate as a function of zenith angle $\theta$
can be described by a $\cos^2 \theta$ dependence
\cite{Beringer:2012aa}. Figure~\ref{fig:zenith_angle} shows a typical
result plot for a scintillator tile distance of $d = 20$~cm and a
measurement time of $1$~h at each zenith angle.

\subsection{Muon Lifetime}

The CosMO detector allows a measurement of the mean lifetime of
muons. Three scintillator boxes are stacked on top of each
other. Decaying muons leave a characteristic signature in one of the
channels, which are two pulses in a time interval, typically in the
microsecond range. The first of these pulses is induced by the
stopping muon, the second one by the electron created in the muon
decay. To suppress coincident muon events, the downmost channel can be
used as a veto.

To enhance the stopping power of the detector, students can test the
influence of different absorber materials between the scintillation
detectors.  A histogram of the recorded time intervals can be fitted
with an exponential function by the use of \textit{muonic}, and the
measured mean lifetime is displayed by the software.  To gather enough
statistics to be confident in the fit, a measurement time of about one
week is required. However, the results of the measurement must be
inspected carefully, since for some scintillators the results are
affected by electronic noise. To suppress this noise contribution,
very short time intervals can be excluded from the fit. The values of
these time intervals can be set via a GUI element within
\textit{muonic}.  The measurement of muon decay gives students
insights into coincidence and veto techniques typically used in
particle physics. This experiment illustrates also the measurement of
statistically fluctuating quantities as well as the law of exponential
decay. The measurement of muon lifetime also provides a good
introduction into discussions about special relativity, because muons
are only able to reach Earth's surface due to relativistic effects.

\begin{figure}[ht!]
  \centering
  \includegraphics[width=0.48\textwidth]{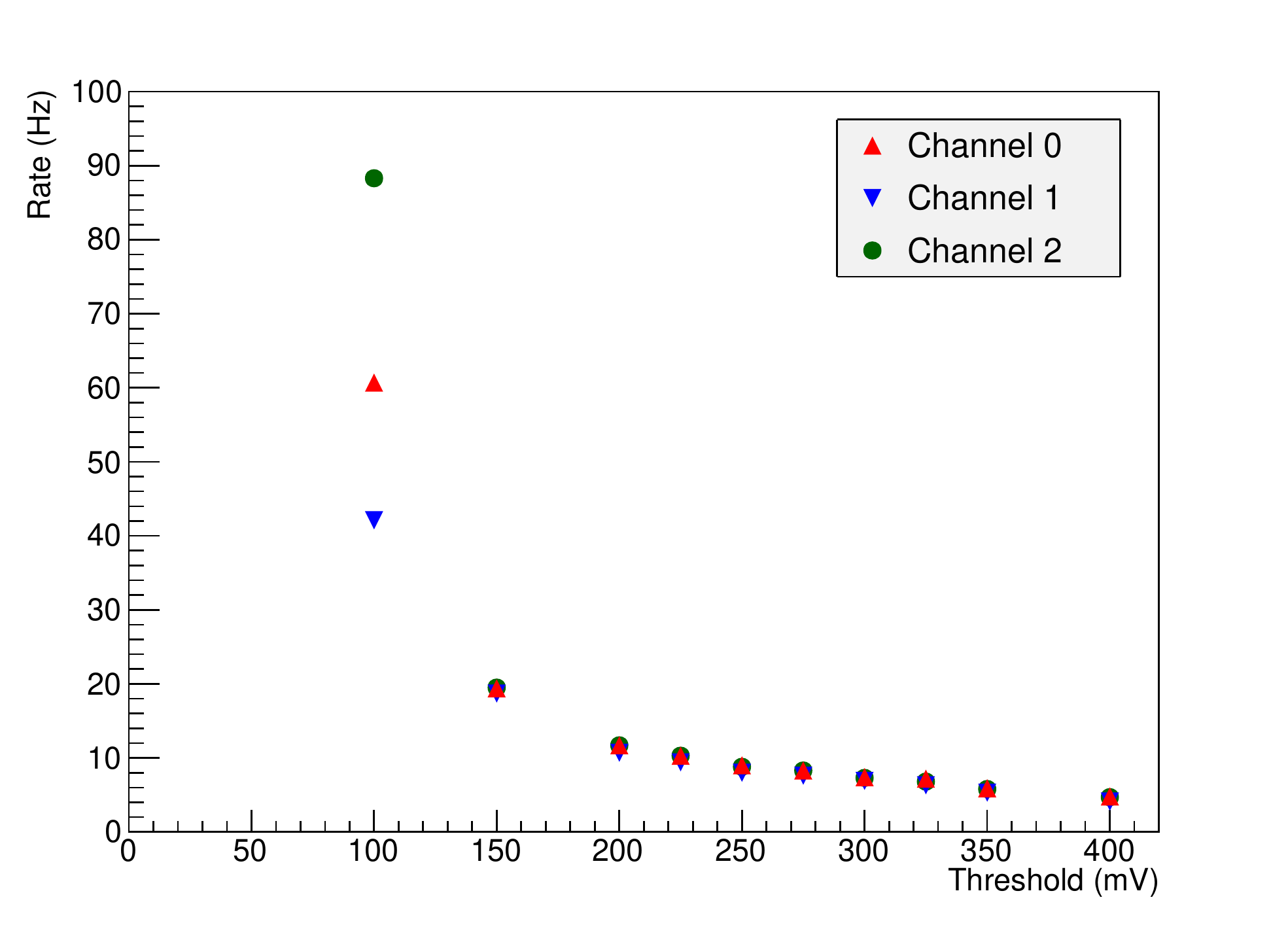}
  \caption{Single channel trigger rate as a function of threshold. The
    operating threshold for each scintillator is chosen on the plateau
    between $250$ and $350$~mV.}
  \label{fig:calibration}
\end{figure}

\subsection{Further Possibilities}

The CosMO detector further allows the study of absorption of
air shower particles in different materials and to measure the
velocity of muons. With three detectors in coincidence and arranged in
a plane, extended particle showers can be measured. The versatility of
the setup enables students to learn about the acceptance of different
detector geometries. With the stored data students can learn about
statistical methods for analysing large data sets.

\section{Netzwerk Teilchenwelt}

\textit{Netzwerk Teilchenwelt} \cite{NTW:URL} is a network of 24
German research institutes of astroparticle and particle physics with
the goal to enable students to authentically experience modern physics
research. More than 100 young scientists are active in the network and
provide students and teachers with insights into their research in
astroparticle and particle physics.  Students and teachers experience
the world of quarks, electrons and cosmic rays firsthand at workshops
in schools, student laboratories, or museums all over Germany. As a
scientist for one day, they analyse real LHC data in a
masterclass. Within cosmic particle projects they perform own
measurements.  \textit{Netzwerk Teilchenwelt} encourages discussion
with scientists and diving into the world of smallest particles and
the big questions about the origin and structure of the universe.

\section{Conclusions and Outlook}

The CosMO experiment enables students and teachers to explore the
physics of cosmic rays in outreach projects or at workshops in
school. They develop their own research project, analyse their own
data, and discuss their results with professional scientists. This
introduces students to the world of the smallest particles and to
questions about the origin and structure of our universe. The hands-on
experiments are currently extended by projects where the scintillation
counters are installed at the German Antarctic research station
Neumayer III \cite{Neumayer:URL} and on the German research icebreaker
Polarstern \cite{Polarstern:URL} of the Alfred-Wegner Institute for
Polar and Marine Research. These detectors take data continuously,
thus enabling analyses over long time ranges with sufficient
statistics. The major goal of the Polarstern project is to measure the
dependence of the rate of cosmic rays on the geographical latitude,
i.e.~the geomagnetic cut-off. Furthermore, the influence of
atmospheric pressure and temperature on the rate of atmospheric muons
can be studied. On the Polarstern and at the Neumayer station
additional neutron monitors are installed. All data will be made
available on the internet to interested students and teachers and can
be analysed together with the data from the CosMO detectors.

All these activities are implemented in and supported by
\textit{Netzwerk Teilchenwelt}. Everybody, no matter whether
scientist, teacher, or student, is encouraged to become active and
join us at \textit{Netzwerk Teilchenwelt}.

\begin{figure}[t!]
  \centering
  \includegraphics[width=0.48\textwidth]{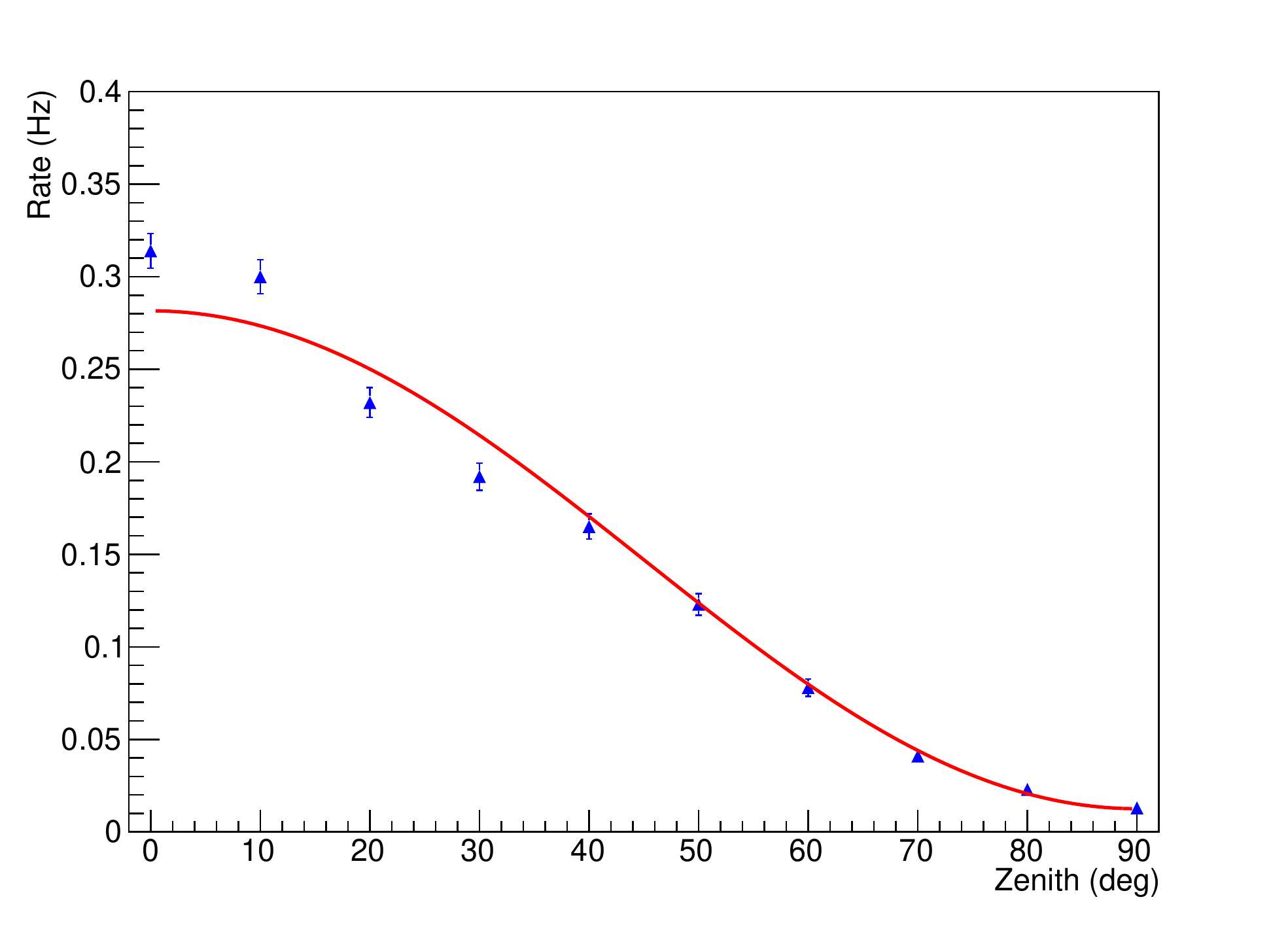}
  \caption{Measured muon rate as a function of zenith angle. To guide
    the eye, the curve indicates the expected \cite{Beringer:2012aa}
    $\cos^2 \theta$ dependence (including an offset for random
    noise).}
  \label{fig:zenith_angle}
\end{figure}

\vspace*{0.5cm} \footnotesize{{\bf Acknowledgment: }{We would like to
    thank the DESY mechanical and electronics workshops for the
    machining and production of the detector components. The authors
    acknowledge the support from Netzwerk Teilchenwelt which is
    managed by the Technical University Dresden under the auspices of
    the German Physical Society (DPG). We also gratefully acknowledge
    the financial support of the German Ministry for Education and
    Research (BMBF).}

\end{document}